\documentstyle[11pt]{article}

\def\mnras{MNRAS}
\def\apj{ApJ}
\def\apjlett{ApJ}
\def\apjsupp{ApJS}
\def\astap{A\&A}

\def\a218{\alpha_{2-18}}
\def\alph20{\alpha_{2-20}}

\def\elldiss{\ell_{\rm diss}}
\def\eps{\epsilon}

\def\Ldiss{L_{\rm diss}}
\def\Lh{L_h}
\def\Ldiss{L_{\rm diss}}
\def\Ls{L_s}

\def\sigmat{\sigma_{\rm T}}

\def\taut{\tau_{\rm T}}
\def\Te{T_{\rm e}}

\def\Tbb{T_{\rm bb}}

 \mathcode`*="002A

\def\prop{\propto}

\newbox\grsign \setbox\grsign=\hbox{$>$} \newdimen\grdimen \grdimen=\ht\grsign
\newbox\simlessbox \newbox\simgreatbox \newbox\simpropbox
\setbox\simgreatbox=\hbox{\raise.5ex\hbox{$>$}\llap
     {\lower.5ex\hbox{$\sim$}}}\ht1=\grdimen\dp1=0pt
\setbox\simlessbox=\hbox{\raise.5ex\hbox{$<$}\llap
     {\lower.5ex\hbox{$\sim$}}}\ht2=\grdimen\dp2=0pt
\setbox\simpropbox=\hbox{\raise.5ex\hbox{$\propto$}\llap
     {\lower.5ex\hbox{$\sim$}}}\ht2=\grdimen\dp2=0pt
\def\simgreat{\mathrel{\copy\simgreatbox}}

\begin{document}
\begin{center}
\vspace{6cm}
\title
{
{\Large ON THE GEOMETRY OF THE X-RAY EMITTING REGION}\\
{\Large IN SEYFERT GALAXIES}
}
\author{
\vspace{0.5cm}
{\sc Boris E. Stern$^{1,2}$, Juri Poutanen$^2$,
Roland Svensson$^2$,}\\
\vspace{0.5cm}
{\sc Marek Sikora$^{2,3}$, and
Mitchell C. Begelman$^{4,5}$} \\
}

\maketitle
\vspace{2cm}
Accepted for publication in Astrophysical Journal Letters
\end{center}

\vspace{1cm}

\begin{enumerate}
\item{ Institute for Nuclear Research, Russian Academy of Sciences
Moscow 117312, Russia,\\
I: stern@inr.msk.su}
\item{ Stockholm Observatory, S-133 36 Saltsj\"obaden, Sweden \\
I: juri@astro.su.se, svensson@astro.su.se}
\item{Nicolaus Copernicus Astronomical Center, Bartycka 18,
00-716 Warsaw, Poland\\
I: sikora@camk.edu.pl}
\item{ JILA, Univ of Colorado, Boulder, CO 80309, USA \\
I: mitch@jila.colorado.edu}
\item{Also at Department of Astrophysical, Planetary, and Atmospheric Sciences,
University of
Colorado, Boulder}
\end{enumerate}

\pagebreak
\baselineskip 12pt

\begin{abstract}
For the first time, detailed radiative transfer calculations of Comptonized
X-ray and $\gamma$-ray radiation in a hot pair plasma above a cold accretion
disk are performed using two independent codes and methods. The simulations
include both energy and pair balance as well as reprocessing of the X- and
$\gamma$-rays by the cold
disk.  We study both plane-parallel coronae as well as active dissipation
regions having shapes of hemispheres and pill boxes located on the disk
surface. It is shown, contrary to earlier claims, that plane-parallel
coronae in pair balance have difficulties in selfconsistently reproducing the
{\it ranges} of 2-20 keV spectral slopes,
high energy cutoffs, and compactnesses inferred from observations of type
1 Seyfert galaxies.  Instead, the observations are consistent
with the X-rays coming from a number of individual active regions located
on the surface of the disk.

A  number of effects such as anisotropic Compton scattering, the reflection
hump, feedback to the soft photon source by reprocessing, and an active
region in pair equilibrium all conspire to produce the observed {\it ranges}
of X-ray slopes, high energy cutoffs, and compactnesses.
The spread in spectral X-ray slopes  can be due to a spread in the properties
of the active regions such as their compactnesses and their elevations
above the disk surface. Simplified models invoking isotropic Comptonization
in spherical clouds are no longer sufficient when interpreting the data.
\end{abstract}

\noindent
{\it Subject headings:} accretion disks --  gamma rays: theory --
radiation mechanism: thermal -- radiative transfer
-- galaxies: Seyfert -- X-rays: general

\pagebreak

\section{Introduction}
\label{sec:intro}

X-ray spectra of Seyfert 1 galaxies in the 2-18 keV range
show an intrinsic power-law component with intensity index
in the range $\alpha=0.9-1.0$ superimposed on a
component arising from reflection (and reprocessing)
of the intrinsic power-law  by cold opaque matter
(e.g. the analysis of 60 spectra from 27 Seyferts by
Nandra \& Pounds 1994).
The overall 2-18 keV spectral index is distributed
around a mean value of  $\a218$ = 0.73 with a dispersion of 0.15.
The reprocessing matter must subtend $\sim$ 1-2 $\pi$ solid angle
as viewed from the X-ray source.
Combined {\it Ginga} and {\it CGRO} OSSE observations show
a spectral cutoff of the intrinsic power-law component
in the several hundred keV range both for IC 4329A
(Zdziarski et al. 1994, Madejski et al. 1995)
and for the average spectrum of a sample of Seyfert 1s
(Zdziarski et al. 1995). NGC 4151 differs in several
respects, lacking a reflection component
and having its cutoff at about 50 keV (Maisack et al. 1993;
Zdziarski, Lightman \& Macio{\l }ek-Nied\'zwiecki 1993),
and thus requires special considerations (Sikora et al. 1995).
A strong annihilation line, predicted by nonthermal models
(e.g. review by Svensson 1994), has never been detected.
The strong UV bumps indicated in many sources
require the existence of thermally radiating cold matter
in the compact region where most of the power is released.

These data are broadly consistent with a geometry where the hot X-ray emitting
gas is located above the cold accretion disk
(as proposed, e.g., by Paczy\'nski 1978). Such a cold disk
is one of the main features
in the canonical black hole models of AGNs.
The data favor {\it thermal} Comptonization of the disk UV radiation
as the main process by which hot electrons
generate the power-law X-rays (Haardt \& Maraschi 1991).
About half of the X-rays impinge on the cold disk
and are reprocessed, emerging mostly as soft black body  radiation,
in addition to any black body radiation
from internal dissipation in the disk.
Haardt \& Maraschi (1991)  emphasized the coupling between the corona and
the cold disk due to reprocessing, as the soft photons affect the cooling
of the hot corona.  In order for the Comptonized spectrum to have a
power-law index close
to unity, nearly all of the power must be dissipated in the hot gas.
Consequently, the soft black body UV luminosity,
$\Ls$, should be of about the same magnitude as the hard Comptonized X-ray
luminosity, $\Lh$, that leaves the corona-disk system.  This is contrary
to observations,
which show that $\Ls$ is often several times larger than $\Lh$.
Another important consequence of the corona-disk geometry (in which the
soft photons enters
the corona from only  one side) is that the source of soft photons is
anisotropic.  To deal with
effects arising from anisotropy, which had been neglected in previous
treatments, Haardt (1993)
developed an approximate theory for anisotropic Compton scattering in a
hot slab. Haardt \&
Maraschi (1993, hereafter HM93) then applied this theory to the corona-disk
geometry, including
reprocessing by the cold disk and imposing pair balance in the hot slab.
 The latter
constraint is crucial, as the coronal temperatures and ``compactnesses"
are sufficiently high that
pair production could account for most of the scattering medium.  HM93
claim that the resulting
X-ray spectral slopes and compactnesses are in excellent agreement with
 observations.

The pair balance calculations by HM93, however, suffer from adopting
the {\it prescription} in Zdziarski (1985) for the
Comptonized spectra including a simple exponential cut off
at photon energies above $kT_e$.  As the dimensionless coronal temperature,
$\Theta \equiv kT_e/ m_e c^2$, is less than unity
in these simulations, only photons in the exponential tail can produce
pairs. Self-consistent
coronal temperatures, $\Theta$, and Thomson scattering optical depths,
$\taut$, are obtained by
first solving the Comptonization and energy balance problem
in the hot gas. However, if the prescribed spectral shapes at
pair-producing energies are erroneous,
one will derive erroneous compactnesses when solving the pair balance
equation for given
$(\Theta, \taut)$.  This
would then lead to incorrect conclusions regarding the spectral properties
as a function of compactness for slab geometry.

In this paper, we use two different and independent
methods of solving exactly the radiative transfer/Comptonization problem
in mildly
relativistic (or relativistic) thermal plasmas, taking into account energy
and pair balance
and cold disk reprocessing (including angular anisotropy and
Klein-Nishina effects).  For a given compactness in {\it slab geometry},
we find temperatures
and spectral slopes which are {\it incompatible} both with HM93's
approximate model and with
observations.  However, models with {\it localized active regions},
simulated by hemispheres or
pillboxes atop a cold disk, can reproduce the observations easily.  The
latter type of geometry
was recognized by Haardt, Maraschi, \& Ghisellini (1994) as a way of
avoiding the rough
equality of $\Ls$ and $\Lh$ that is inevitable in uniform slab models.

The first method used in our calculations is based on the Nonlinear Monte
Carlo (NLMC)
method developed by Stern (1985, 1988) and described in detail in Stern et al.
(1995). This method can follow particles and photons
interacting with a background medium consisting of the
same particles and photons (this is the non-linearity)
in both steady and time-dependent systems with any chosen geometry.
In the present case, only photons are followed while the electrons and
positrons
are assumed to be thermal.  The second method is a
pure 1D radiative transfer code for Compton scattering
in relativistic plasmas developed by
Poutanen (1994; see also Poutanen \& Vilhu 1993) using the iterative
scattering method (ISM),
where the radiative transfer equation is solved for each scattering
order (e.g. Sunyaev \& Titarchuk 1985, Haardt 1994).
A thorough discussion of the latter ISM code
and a comparison with results obtained using the NLMC code will appear in
Poutanen et al. (1995).

In \S~\ref{sec:slab} we apply these methods to coronal slabs,
and in \S~\ref{sec:clouds} to two geometries for active regions finding
that the results for active regions
are more consistent with observations.

\section{Comptonization in Plane-Parallel Pair Coronae (Slabs) }
\label{sec:slab}

\subsection{Setup}
\label{sec:slabsetup}

In both types of simulation, a local dissipation compactness,
$\elldiss \equiv ( \Ldiss /h ) (\sigmat /  m_e c^3)$,
characterizes the dissipation with $\Ldiss$ being the power providing uniform
heating in a cubic volume of size $h$ in the slab, where $h$ is the height
of the
slab. For the purpose of obtaining accurate radiative transfer,
the slab is divided into $3-11$ homogeneous spatial
zones depending upon the expected $\taut$.
The reprocessed radiation has a flux equal to the absorbed incident flux and a
Planckian spectral {\it shape}, but
for simplicity the Planckian temperature, $\Tbb$ is set to a value
of 5 eV, a typical black body temperature
for canonical accretion disks in AGNs.
We assume no internal dissipation in the cold disk.
The two parameters, $\elldiss$ and $\Tbb$, uniquely specify the slab
simulations. We consider pure pair coronae, assuming that any coronal
background plasma has a scattering optical depth much less than $\taut$.
For both types of codes, the system is relaxed
towards a steady state as regards energy balance, pair balance,
and emerging spectra. On a Sun IPX, a typical NLMC simulation takes
a few hours, while the ISM code takes about 10 minutes for
6 angular gridpoints, 7 spatial zones, and 80 frequency points.

\subsection{The Energy and Pair Balance in Slabs}
\label{sec:slabbalance}

In the NLMC simulations, the spatial distributions of
temperature and pair density are almost homogeneous,
having a contrast of less than a factor of 2 between the top and bottom
slab layers
as a result of $\taut$ being less than unity.
As output parameters, the temperature, $T_e$ (or $\Theta$),
is averaged over height, and the vertical Thomson scattering optical depth,
$\taut$, is easily computed. With the ISM code, the
energy balance and pair balance are solved using
height-averaged Compton cooling and pair production, giving
a typical temperature and pair density for the almost homogeneous slab.
The {\it rectangles} and the {\it dashed curve} in Figure 1a
show $\Theta$ vs. $\elldiss$ obtained
with the NLMC code and the ISM code, respectively.
Figure 1b shows $\taut$ vs. $\elldiss$ using the same notation.
The agreement is very good.

Formally, the problem at hand can be solved in three steps.
First, the radiative transfer/Comptonization problem together with
energy conservation for the cold disk (i.e. reprocessed flux equals incident
flux)
is solved giving
a relation between $\Theta$ and $\taut$.
As expected, the $\Theta$ vs. $\taut$ relation obtained with the approximate
treatment of HM93 agrees with our exact simulations to within a few \%
up to our largest value of $\taut = 0.37$.  The solutions
are characterized by an almost constant generalized Kompaneets parameter,
$y \equiv \taut (1 + \taut) (4\Theta + 16 \Theta^2)$.
For the four NLMC solutions, $y \approx 0.48$, while for the ISM solutions,
$y \approx 0.49-0.53$.
Secondly, the energy balance of the corona is solved
giving the {\it ratio} $\Ldiss/\Ls$, or, alternatively, $\Lh/\Ls$
(where typically $\Lh \sim \Ldiss/2$). In the coronal energy balance,
we account for the reduction of $\Ls$ due to scattering of the
reprocessed spectrum in the slab (important for $\taut \simgreat 0.1$).
For example,
for $\taut$ = 0.37 we find $\Lh/\Ls = 1.60$, while neglecting the
effect (as in HM93) gives $\Lh/\Ls = 0.88$.
Thirdly, the pair balance problem is solved
for previously obtained combinations
of $\Theta$ and $\taut$, giving the absolute value of $\elldiss$
($= 10^3$ for $\taut$ = 0.37) .
We find that the corresponding values of $\elldiss$
obtained by HM93 ({\it dotted curve} in Fig. 1)
are in error by up to a factor of 20 due to the
spectrum prescribed by HM93 at photon energies, $h\nu > kT_e$.

\subsection{Spectra from Slabs}
\label{sec:slabspectra}

For improved statistics with the NLMC method, the spectra are averaged over
viewing angles $0.6 < \cos \theta < 1$, i.e. approximately the viewing angles
expected for Seyfert 1s in the unified model. Here, $\theta$ is the angle
relative to the normal of the disk. With the ISM  code,
spectra are computed at the chosen angular gridpoints
($\cos \theta =$ 0.113, 0.5, 0.887) in the upward direction.
The least square overall spectral slope, $\a218$, for the
2-18 keV range were determined and are displayed
in Figure 2 by the rectangles (using the NLMC code) and by the
{\it dashed curve} (using the ISM code for $\cos \theta$ = 0.887).
For $\elldiss > 1$ we find that $\a218$ at $\cos \theta \sim 0.9$ (i.e.
almost face-on) and 0.5 differ by less than 5 \% allowing comparisons
between the face-on results using the ISM code and the average spectra from
the NLMC code. The two codes give spectra that are in very good agreement,
providing
the first serious test of both codes.
The right panel of Figure 2 shows the observed distribution of
$\a218$ for {\it Ginga} spectra from 27 Seyfert galaxies (Nandra \& Pounds
1994). We conclude from Figure 2 that slabs must have $\elldiss < 1$
in order to reproduce the observed range of $\a218$, which is in conflict
with the best available (albeit uncertain) observed compactnesses
($\ell_h \sim \elldiss/2 \sim 1-100$, e.g. Done \& Fabian 1989).

We find our calculated spectra to be in
qualitative, but not quantitative, agreement with the spectra in Figures
4a-c in HM93.
The true spectral shape at $h\nu > k\Te$ in
an optically thin plasma has a slow cutoff reflecting the thermal
Compton scattering kernel (approximately $\prop \exp(-h\nu/2k\Te)$) rather
than the more
rapid ad hoc exponential cutoff, $\nu^{1-\alpha} \exp(-h\nu/k\Te)$, of HM93.
Having too few pair-producing photons,
HM93 obtained $\elldiss$ (they used notation $\ell_{\rm c}$)
up to a factor of 20 too large when solving the pair balance equation.
Therefore, HM93 erroneously found slab models to be in agreement with
observed spectra and compactnesses.

\section{Comptonization in Active Pair Regions}
\label{sec:clouds}

\subsection{Setup}
\label{sec:cloudsetup}

We represent the active regions in different ways with the two codes.
With the NLMC code that fully handles 2D and 3D radiative transfer,
we consider hemispheres of radius $h$ located on the disk surface,
and divided into 5 homogeneous spatial zones as shown in Figure 7
in Stern et al. (1995). The hemisphere problem is axisymmetric and thus 2D.
As the ISM code can only handle 1D problems, cylinders of
pill box shape of height $h$ and diameter $2h$ were considered.
By allowing reprocessed and reflected radiation to enter only at the bottom
of the pill box, assuming a non-zero source function only within the pill box,
and averaging the computed radiation field over each spatial layer  for
each scattering  order, we effectively convert the problem to 1D.
We also take account of the fact that not all of the reprocessed
X-ray radiation reenters the active region (about 60 \% for the hemisphere,
see also Haardt 1994; about 50 \% for the pill box).
As for the slab case, these simulations depend on
two parameters, $\elldiss$ and $\Tbb$, but now $\Ldiss$ is the power dissipated
uniformly in the active region. As before, $\Tbb$ = 5 eV.

\subsection{The Energy and Pair Balance in Active Pair Regions}
\label{sec:cloudbalance}

In the output of our NLMC simulations,
the computed temperature, $T_e$ (or $\Theta$),
is averaged over volume, and a mean Thomson scattering optical depth,
$\taut$, is obtained by averaging over all radial directions of the hemisphere.
With the ISM code, volume-averaged
energy balance and pair balance are solved for
a typical $\Theta$ and the vertical $\taut$ of the almost homogeneous pill box.
The {\it hemispheres} and {\it solid curves} in Figure 1a
show $\Theta$ vs. $\elldiss$ obtained
with the NLMC code and the ISM code, respectively.
Figure 1b shows $\taut$ vs. $\elldiss$ using the same notation.
The parameter combinations ($\taut$, $\Theta$) obtained for different
$\elldiss$ give almost constant Kompaneets parameter, $y$.
We find $y \approx 1.9-2.2$ for hemispheres, and $y \approx 2.0-2.4$ for
pill boxes.
The reason for the larger $y$ for active regions,
as compared to slabs,
is that fewer of the soft reprocessed
photons return to the active region (yielding increased ``photon starvation"
e.g. Zdziarski, Coppi, \& Lamb 1990),  thus requiring harder spectra to
achieve the needed Compton cooling, and correspondingly larger values of $y$.

At given $\elldiss$, values of $\Theta$ are
larger for active pair regions as compared to pair slabs,
while $\taut$ is smaller at small $\elldiss$ and larger at large $\elldiss$.
The closer the active pair regions are, the more they will influence each other
by providing X-ray photons for Comptonization and $> 511$ keV photons
for pair production. Such ``interacting"
pair regions will have $\Theta$ and $\taut$ in between the curves
for slabs and active regions in Figure 1a. A slab corona is the limit
of closely packed active regions.
Active pair regions near disks have properties placing
them in the parameter space in Figure 1a forbidden for pair slabs.

\subsection{Spectra from Active Pair Regions}
\label{sec:cloudspectra}

Figure 3 shows emerging spectra from hemispheres
at different $\elldiss$.
Each {\it solid curve} shows a total spectrum, being the sum of
the Comptonized spectrum from the hemisphere ({\it dotted curve}),
the reprocessed black body spectrum at 5 eV, and the reflection component.
In mildly relativistic plasmas, Compton scattering in the
forward direction is suppressed, causing
the contribution of once scattered photons in face-on directions to
be suppressed (HM93). This effect of anisotropy causes
the Comptonized spectra ({\it dotted curves}) to be broken
power-laws with the {\it anisotropy break}
occurring just above the peak energy of
twice scattered photons. The power-law slope at energies
above the break is related to the Kompaneets parameter, $y$.
The fact that $y$ is almost independent of $\elldiss$ shows through the
almost constant
spectral slope of the Comptonized spectra above the break.
As $\elldiss$ increases and $\Theta$ decreases the break moves to lower
energies
through the 2-18 keV range causing $\a218$ to soften from 0.55 to 1.
For viewing angles relevant to Seyfert 1s,
the supression of once-scattered photons does not affect
the spectra in the 2-18 keV range except at small $\elldiss$
when this Compton order extends into that spectral range (see Fig. 3 for case
$\elldiss = 3$). The angle-averaging procedure we use washes out
some of the anisotropy at lower energies, but produces the correct $\a218$
as there is little spread in $\a218$ for the considered range
of viewing angles.

The spectral slopes, $\a218$, are shown in Figure 2 for both {\it hemispheres}
and pill boxes ({\it solid curve}) as a function of $\elldiss$.
The right panel shows the observed distribution of
$\a218$ for {\it Ginga} spectra from 27 Seyfert galaxies (Nandra \& Pounds
1994). Active regions produce spectra covering
the {\it observed ranges} of $\a218$ ($\approx 0.4-0.9$)  and cutoff
energies ($\sim 2kT_e$)
for the {\it observed range} of compactnesses
($\ell_h \sim \elldiss/2 \sim 1-100$, Done \& Fabian 1989).
The observed mean $\a218\sim 0.7$ implies a preferred value of $\elldiss
\sim 20$.
Elevating the active region from the disk would make even fewer soft
reprocessed photons return to the active region, requiring larger $y$ and thus
harder spectra. The corresponding {\it solid} curve would thus be lowered
in Figure 2 by an amount depending upon the elevation.
The observed dispersion of $\a218$ could be due both to
a spread in $\elldiss$ and in the elevation of active regions above the disk.

\section{Summary}
\label{sec:summary}

Our results can be summarized as follows:
(1) We have performed the first exact radiative transfer simulations
of Comptonization by hot gas at mildly relativistic temperatures located
in the vicinity of a cold disk, fully accounting for geometric effects,
reprocessing, and self-regulation of the soft photon flux,
pair and energy balance. We have used two codes which rely on quite different
methods,
and have obtained excellent agreement for cases where detailed comparison
can be made.
(2) We find that pair slab models with  $\Tbb$ = 5 eV have difficulties in
reproducing
the observed characteristics of type 1 Seyfert X-ray spectra, while
active pair regions with geometries of, e.g., hemispheres or pill
boxes produce X-ray spectra in much better agreement with observations.
Whether this is still true for larger $\Tbb$ remains to be seen. The dispersion
of observed spectral behavior is likely due to the
active regions having a (nonuniform) spread in compactness and elevation
above the disk.
(3) Our findings support the idea of a structured or patchy corona
with a number of separated active regions (e.g. flares and coronal loops)
first suggested by Galeev, Rosner, \& Vaiana (1979), and recently rejuvenated
by Haardt, Maraschi, \& Ghisellini (1994) who argued
that active regions typically would have $\elldiss \sim 30$ in agreement with
our preferred $\elldiss$.
(4) We finally conclude that simplified models neglecting effects of anisotropy
and other geometric effects,
reprocessing, and correct spectral behavior at pair-producing energies
are no longer adequate when interpreting spectra.

{\bf Acknowledgments:} This research was partially supported by
NASA grants NAG5-2026, and NAGW-3016;
NSF grants AST91-20599 and INT90-17207; the Polish KBN grant
2P03D01008; and by grants from the Swedish Institute and the Swedish
Natural Science Research
Council. The authors thank Francesco Haardt for insightful comments.

\clearpage

\begin{center}
\Large{\bf Figure Captions}
\end{center}

FIG.~1--- (a) Dimensionless temperature, $\Theta \equiv kT_e/ m_e c^2$,
vs. dissipation compactness,
$\elldiss \equiv ( \Ldiss /h ) (\sigmat /  m_e c^3)$,
for
a steady X-ray emitting region in pair and energy balance located on
a cold disk surface. The soft photons from the cold disk are assumed to
have $k\Tbb$ = 5 eV.
{\it Rectangles} and {\it dashed curve} show results
from NLMC code and ISM code, respectively, for
the case of a plane-parallel slab corona.
{\it Dotted curve} shows results for slabs from the approximate treatment
of HM93. Results for individual active pair regions
located on the disk surface are shown by
{\it hemispheres} (hemisphere geometry using NLMC code) and {\it solid curve}
(pill box geometry using ISM code).
The parameter space to the right of respective curves is
forbidden as pair balance cannot be achieved, and  the parameter space
to the left would contain solutions where the background
coronal plasma dominates over the pairs, i.e. ``pair free" solutions
(e.g. Svensson 1984, HM93).
(b) Vertical Thomson scattering optical
depth, $\taut$, vs. dissipation compactness,
$\elldiss$. Same notation and parameters as in (a).

FIG.~2--- Overall spectral intensity index, $\a218$, least square fitted
to the model spectra in the 2-18 keV range vs. the dissipation compactness,
$\elldiss$. Same notation as in Fig. 1.
The spectra from NLMC code were averaged over viewing  angles
$0.6 < \cos \theta < 1.0$ before determining $\a218$.
For the ISM code, face-on spectra (at $\cos \theta = 0.887$) were used.
The right panel shows the observed distribution of
$\a218$ for {\it Ginga} spectra from 27 Seyfert galaxies (Nandra \& Pounds
1994).

FIG.~3---  Emerging spectra, $\eps F_{\eps}$, where $F_{\eps}$ is the energy
flux (arbitrary units) and $\eps \equiv h \nu / m_e c^2$ from hemispheres
of different compactnesses, $\elldiss$.
The spectra are averaged over
viewing angles $0.6 < \cos \theta < 1$.
The {\it  solid curves} show the total spectrum which is the
sum of the Comptonized spectrum from the hemisphere itself
({\it dotted curves}),
the reprocessed black body spectrum, and the reflection component.
Vertical {\it dashed lines} show the 2-18 keV spectral range.


\clearpage


\begin{thebibliography}{}
\bibitem{don89}
 Done, C., \& Fabian, A. C. 1989, \mnras, 240, 81
\bibitem{gal79}
 Galeev, A.A., Rosner, R., \& Vaiana, G. S. 1979, \apj, 229, 318
\bibitem{haa93}
Haardt, F. 1993, \apj, 413, 680
\bibitem{haa94}
 Haardt, F. 1994, PhD thesis, SISSA, Trieste
\bibitem{hm91}
 Haardt, F.,\& Maraschi, L. 1991,  \apjlett, 380, L51
\bibitem{hm93}
Haardt, F.,\& Maraschi, L. 1993, \apj, 413, 507 (HM93)
\bibitem{hmg94}
 Haardt, F., Maraschi, L., \& Ghisellini, G. 1994, \apjlett, 432, L95
\bibitem{mad95}
 Madejski, G. M. et al. 1995, \apj, 438, 672
\bibitem{1993}
Maisack, M., et al.\ 1993, \apjlett, 407, L61
\bibitem{nan94}
 Nandra, K., \& Pounds, K. A. 1994, \mnras, 268, 405
\bibitem{pac78}
 Paczy\'nski, B. 1978, Acta Astr., 28, 241
\bibitem{pout95}
Poutanen, J., et al. 1995, in preparation
\bibitem{pout94}
 Poutanen, J. 1994, PhD thesis, University of Helsinki
\bibitem{pou93}
 Poutanen, J. \& Vilhu, O. 1993, \astap, 275, 337
\bibitem{sik95}
 Sikora, M. et al. 1995, in preparation
\bibitem{ste85}
Stern, B. E. 1985, Sov. Astr., 29, 306
\bibitem{ste88}
 Stern, B. E. 1988, Nordita/88-51 A, preprint
\bibitem{ste95a}
 Stern, B. E., Begelman, M. C., Sikora, M., \& Svensson, R. 1995,
         \mnras, 272, 291
\bibitem{sun85}
 Sunyaev, R. A. \& Titarchuk, L. G. 1985, \astap, 143, 374
\bibitem{sve84}
 Svensson, R. 1984, \mnras, 209, 175
\bibitem{sve94}
 Svensson, R. 1994, \apjsupp, 92, 585
\bibitem{zdz85}
 Zdziarski, A. A. 1985, \apj, 289, 514
\bibitem{zdz94}
 Zdziarski, A. A. et al. 1994, \mnras, 269, L55
\bibitem{zdz95}
 Zdziarski, A. A., Johnson, W. N., Done, C., Smith, D.,
           \& McNaron-Brown, K. 1995, \apjlett, 438, L63
\bibitem{zdz90}
 Zdziarski, A. A., Coppi, P. S., \& Lamb, D. Q. 1990, \apj, 357, 149.
\bibitem{zdz93}
 Zdziarski, A. A., Lightman, A. P., \& Macio\l ek-Nied\'zwiecki, A.
           1993, \apjlett, 414, L93
\end{thebibliography}
\end{document}